\documentclass[aps,prb,twocolumn,floats,showpacs]{revtex4}

\usepackage{epsfig}

\usepackage{amsmath}
\usepackage[dvips]{color}
\usepackage{amssymb}

\usepackage{bm}

\begin{document}

\title{Kondo Tunneling into a Quantum Spin Hall Insulator}

\author{Igor Kuzmenko, Anatoly Golub and Yshai Avishai}
\affiliation{Department of Physics, Ben-Gurion University of
  the Negev Beer-Sheva, Israel}
\date{\today}

\begin{abstract}
The physics of a junction composed of a normal metal, quantum dot
and 2D topological insulator (in a quantum spin Hall
state) is elucidated. It maifests a subtle combination of Kondo
correlations and quantum spin Hall  edge states moving on the
opposite sides of the 2D topological insulator. 
In a narrow strip geometry these edge states interact and a gap opens
in the edge state spectrum.  Consequently, Kondo screening is 
less effective and that affects
electron transport through the junction. Specifically, when
edge state coupling is strong enough, the tunneling differential
conductance develops a dip at zero temperature instead of the
standard  zero bias Kondo peak.
\end{abstract}


\maketitle

\noindent
{\bf Background:} Topological insulators (TI) and topological
superconductors (TS) have recently become a
subject of intense theoretical and experimental studies.
\cite{Teo-Kane-10, 3DTI-08, TI-and-TS-09, Kane-Mele-prl03, %
Liang-Kane-Mele-prl07, Hasan-Kane-rmp10, Liang-Kane-prb07, %
Liang-Kane-prl08, Bi2Te3-and-Sb2Te3-prl09, Bi2Te3-prl09, %
Bi2Se3-prb09, HgMnTe-TI-08, TI-10, Bi2Se3-photo-prl09} In most of
these studies, properties of topological states are analyzed using
classification of the electronic phases according to the pertinent
topological invariants.\cite{Teo-Kane-10, 3DTI-08, TI-and-TS-09, %
Kane-Mele-prl03, Liang-Kane-Mele-prl07, Hasan-Kane-rmp10, %
Liang-Kane-prb07, Liang-Kane-prl08}

Two-dimensional TI are systems in which time reversal symmetry is
respected but spin rotation invariance is violated. They differ
from ordinary insulators as they posses a non-trivial topological
number $Z_2$ implying a gapped spectrum in the bulk and gapless
spectrum of quantum spin Hall (QSH) edge states\cite{konig} (or
helical modes). A helical mode consists of a Kramers degenerate
pair of states propagating in opposite directions along the same
edge. Elucidating the QSH helical modes in experiments is rather
elusive since both edge states moving on opposite boundaries
contribute to the conductance.\cite{egger} This obstacle can be
alleviated if there are strong correlations of edge states on the
opposite edges, leading to the emergence of massive Dirac fermions
instead of the massless ones that prevail in the absence of such
correlations.

\noindent
{\bf Motivation:} In order to employ this feature, we note that
the physics of QSH state can be studied in tunneling experiments.
\cite{Tkachev-11, TI-trans-09, TI-trans-10, TI-trans-11, GKY-11, %
ScanTunMicr-prl11,TI-tun-10} In particular, we focus on tunneling through
an interacting quantum dot tuned to the Kondo regime.
In the absence of correlations between edge states,
Kondo tunneling  into a TI with gapless
helical edge state spectrum displays a finite zero bias differential
conductance at zero
temperature\cite{TI-tun-10}, as is the case for  tunneling into a
 metal. Once the coupling between the helical states on the
opposite sides of the TI is present, a gap opens in the edge states
spectrum, which naturally affects
the tunneling conductance.\cite{egger} Screening of the
magnetic impurity becomes less effective and
the conductance develops a dip at zero temperature.

\noindent
{\bf The main objective:} The main goal of the present work is to substantiate
the above qualitative analysis. It is argued that
interaction between QSH edge states on a 2D TI has a clear signature
that turn it to be experimentally detectable as
it leads to unusual behavior of
Kondo tunneling at zero temperature.

\begin{figure}[htb]
\centering
\includegraphics[width=80 mm,angle=0]
   {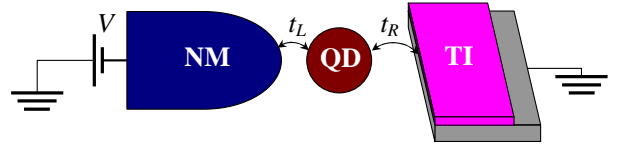}
 \caption{NM-QD-TI junction. Helical states exist on the edges
  of the TI. The QD can be gated to adjust the level position in
  the dot and/or the tunneling rates $t_L$ and $t_R$.}
  \label{Fig-NM-QD-TI}
\end{figure}

\noindent
{\bf Main results:} We have theoretically analyzed tunneling through a device
consisting of a tunnel junction composed of normal metal (NM)
interacting quantum dot (QD) and 2D TI as shown in Fig.
\ref{Fig-NM-QD-TI}. The TI is modeled as a stripe with topological
states moving on its opposite sides with a coupling between
them.  Interrelation between the
Kondo physics prevailing in the QD and the QSH physics prevailing
in the TI is then studied. Kondo tunneling is analyzed in the weak ($T\gg{T_K}$)
and strong ($T<T_K$) coupling regimes, where $T_K$ is a Kondo
temperature. It is shown that in both regimes, the conductance
develops a dip when the temperature $T$ decreases and crosses an
energy threshold $\nu$  specifying the coupling strength between
the QSH edge states. Since this setup is experimentally feasible we argue that
coupling between QSH edge states residing on opposite edges of a 2D TI
has a definite experimental signature.


\noindent
{\bf Anderson Model for NM-QD-TI Junction:}
The junction consists of the following ingredients (see Fig.
\ref{Fig-NM-QD-TI}): the NM lead, held at bias
voltage $V$; the TI with coupling between
helical modes on its two opposite sides; the QD
 in a tunnel contact with the NM and with the TI. The Hamiltonian
of the junction is then written as,
\begin{eqnarray}
  H &=&
  H_{L}+H_{R}+H_{D}+H_T.
  \label{H-Anderson-def}
\end{eqnarray}
The first term on the right hand side of Eq.
(\ref{H-Anderson-def}) is the Hamiltonian of the NM lead,
\begin{eqnarray}
  H_L &=&
  \sum_{{\bf{k}}\sigma}
  \epsilon_{{\bf{k}}}
  c_{{\bf{k}}\sigma}^{\dag}
  c_{{\bf{k}}\sigma}.
  \label{H-lead-def}
\end{eqnarray}
The second term, $H_{R}$, is the low energy Hamiltonian of the
2D TI,
\begin{eqnarray}
  H_{R} =
  \hbar
  \sum_{k\sigma r}
  v_{\sigma r}k~
  \alpha_{k\sigma r}^{\dag}
  \alpha_{k\sigma r}+
  \nu
  \sum_{k\sigma}
  \Big[
      \alpha_{k\sigma1}^{\dag}
      \alpha_{k\sigma2}+
      {\rm{h.c.}}
  \Big],
  \label{H-topins-def}
\end{eqnarray}
where $\nu$ is the coupling strength between the helical fermions on the
opposite sides ($r=1,2$) of the TI, and $v_{\sigma r}\equiv(-1)^{r-1}
\sigma v$, with $v$ being the Fermi velocity.

The coupling $\nu$ represents the effective mass of Dirac
fermions. It is convenient to work in a basis
for which the TI Hamiltonian (\ref{H-topins-def}) is
diagonal,  obtained through the canonical transformations,
\begin{eqnarray}
  &&
  \begin{array}{rcl}
    \alpha_{k\sigma 1}
    &=&
    U_k \gamma_{k\sigma +}-
    \sigma V_k \gamma_{k\sigma -},
    \\
    \alpha_{k\sigma 2}
    &=&
    \sigma V_k \gamma_{k\sigma +}+
    U_k \gamma_{k\sigma -},
  \end{array}
  \label{UV-transform}
\end{eqnarray}
where
\begin{eqnarray*}
  &&
  U_k =
  \sqrt{\frac{\varepsilon_k+\hbar v k}
             {2\varepsilon_k}},
  \ \ \ \ \
  V_k =
  \sqrt{\frac{\varepsilon_k-\hbar v k}
             {2\varepsilon_k}},
  \\
  &&
  \varepsilon_k=\sqrt{(\hbar v k)^2+\nu^2}.
\end{eqnarray*}
The Hamiltonian (\ref{H-topins-def})
then takes the form,
\begin{eqnarray}
  H_{R} &=&
  \sum_{k\sigma, p=\pm}
  p\sigma \varepsilon_k
  \gamma_{k\sigma p}^{\dag}
  \gamma_{k\sigma p}.
  \label{H-topins-diag}
\end{eqnarray}
The third term on the right hand side of Eq.
(\ref{H-Anderson-def}), $H_D$, is the QD Hamiltonian,
\begin{eqnarray}
  H_D &=&
  \epsilon_0
  \sum_{\sigma}
  d_{\sigma}^{\dag}
  d_{\sigma}+U n_{\downarrow}n_{\uparrow}.
  \label{H-dot-Habbard}
\end{eqnarray}
Here $d^{\dagger}_{\sigma},d_{\sigma}$ are creation and
annihilation operators for the dot electrons;
$n_{\sigma}=d^{\dagger}_{\sigma}d_{\sigma}$, and $U>0$ is the strength of the Coulomb repulsion.
The last term on the right hand side of Eq.
(\ref{H-Anderson-def}), $H_T=H_{TL}+H_{TR}$,  is composed of $H_{TL}$, describing
electron tunneling between the dot and the normal metal lead, and $H_{TR}$ describing electron tunneling between the dot and the TI. 

Denoting
by $\psi_L({\bf r})$ and $\psi_R({\bf r})$ the electron field
operators in the NM and the TI and  assuming the dot is positioned at
${\bf r}=0$, we have,
\begin{eqnarray}
&&  H_{TL} =
  t_{L}
  \sum_{\sigma}
  \Big[
      \psi_{L \sigma}^{\dag}(0)
      d_{\sigma}+
      {\rm{h.c.}}
  \Big], \nonumber \\
 && H_{TR} =
  t_{R}
  \sum_{\sigma}
  \Big[
      \psi_{R \sigma}^{\dag}(0)
      d_{\sigma}+
      {\rm{h.c.}}
  \Big],
  \label{H-TLR-def}
\end{eqnarray}
where $t_{L,R}$ are the corresponding tunneling amplitudes assumed
to be spin independent.

\noindent
{\bf The Spin Hamiltonian:}
The quantum dot is tuned by gate voltage
to have a single electron in its ground state.
The depth $|\epsilon_0|$ of the dot level is assumed to be much larger than the tunneling width,
\begin{equation} \label{Gamma}
  \Gamma=\Gamma_L+\Gamma_R,
  \ \ \
  \Gamma_L=
  2\pi t_L^2\rho_L,
  \ \ \
  \Gamma_R=
  2\pi t_R^2\rho_R,
  \end{equation}
where $\rho_L$ is the electron density of states in the NM, while
$\rho_R$ is the electron density of states belonging to the
gapless part of the TI.
Hence, charge fluctuations can be integrated out using  the
%
Schrieffer-Wolff transformation \cite{Schrieffer-Wolff-66}.
It projects out zero and two electron states
in the dot and maps the Hamiltonian $H$, Eq.
(\ref{H-Anderson-def}) onto an effective spin Hamiltonian
$\tilde{H}$ acting in a subspace of states where there is one and only one electron on the dot. The
effective Hamiltonian in this subspace reads,
\begin{eqnarray}
  H_K &=&
  \sum_{\alpha\alpha'\sigma\sigma'}
  \psi_{\alpha\sigma}^{\dag}(0)~
  h_{\alpha\alpha'}^{\sigma\sigma'}~
  \psi_{\alpha'\sigma'}(0),
  \label{H-spin}
\end{eqnarray}
where $\alpha,\alpha'=L,R$, and the matrix $h_{\alpha\alpha'}^{\sigma\sigma}$ is defined as,
\begin{eqnarray*}
  &&
  h_{\alpha\alpha'}^{\sigma\sigma'}=
  \frac{1}{4}~
  K_{\alpha\alpha'}
  \delta_{\sigma\sigma'}+
  \frac{1}{2}~
  J_{\alpha\alpha'}
  {\bf{S}}
  \cdot
  {\boldsymbol\tau}_{\sigma\sigma'}.
\end{eqnarray*}
Here $\boldsymbol\tau$ is the 
vector of the Pauli matrices and ${\bf{S}}$ is the spin operator of
the electron residing on the QD. The coupling constants are
\begin{eqnarray*}
  J_{\alpha\alpha'} =
  \frac{2t_{\alpha}t_{\alpha'}}{|\epsilon_0|}+
  \frac{2t_{\alpha}t_{\alpha'}}{U-|\epsilon_0|},
  &&
  K_{\alpha\alpha'} =
  \frac{2t_{\alpha}t_{\alpha'}}{|\epsilon_0|}-
  \frac{2t_{\alpha}t_{\alpha'}}{U-|\epsilon_0|}.
\end{eqnarray*}

\noindent
{\bf Scaling Equations} Using the poor man's scaling technique, we obtain scaling
equations for the dimensionless coupling constants
$j_{\alpha\alpha'}=J_{\alpha\alpha'}\sqrt{\rho_{\alpha}\rho_{\alpha'}}$,
\begin{eqnarray}
  \frac{d j_{LL}}{d\ln{D}}
  &=&
  -\big(
      j_{LL}^2+
      j_{LR}^2
  \big),
  \label{eq-jLL}
  \\
  \frac{d j_{RR}}{d\ln{D}}
  &=&
  -\big(
      j_{RR}^2+
      j_{LR}^2
  \big),
  \label{eq-jRR}
  \\
  \frac{d j_{LR}}{d\ln{D}} &=&
  -j_{LR}~
  \big(
      j_{LL}+
      j_{RR}
  \big),
  \label{eq-jLR}
\end{eqnarray}
subject to the initial conditions,
\begin{eqnarray*}
  j_{\alpha\alpha'}(\bar{D})=
  \frac{\sqrt{\Gamma_{\alpha}\Gamma_{\alpha'}}}{\pi}
  \left(
       \frac{1}{|\epsilon_0|}+
       \frac{1}{U-|\epsilon_0|}
  \right).
\end{eqnarray*}
If $\nu$ is the smallest energy scale, the solutions of these
equations are,
\begin{eqnarray}
  j_{\alpha\alpha'}(T) &=&
  \frac{\sqrt{\Gamma_{\alpha}\Gamma_{\alpha'}}}
       {\Gamma}~
  \frac{1}{\ln\left(T/T_K\right)},
  \label{j(T)-solution}
\end{eqnarray}
and the scaling invariant, (the Kondo temperature), is
\begin{eqnarray}
  &&
  T_K =
  \bar{D}
  \exp
  \bigg\{
       -\frac{\pi|\epsilon_0|}{\Gamma}
  \bigg\}.
  \label{TK}
\end{eqnarray}

When $\nu$ is not the smallest energy scale, the scaling behavior is more complicated. The
flow diagram still has a fixed point at infinity, but the Kondo
temperature turns out to be a sharp function of $\nu$.
\cite{KikoinAvishai02, PustGlazm00} For
$\nu\gg{T}_K$, the scaling of $j_{\alpha\alpha'}(T)$ depends
on whether the temperature is higher or lower than $\nu$. For $T\gg\nu \gg T_K$, the gap in the edge state spectrum 
can be neglected and the scaling of $j_{\alpha\alpha'}(T)$ are given by
Eq. (\ref{j(T)-solution}) with the scaling invariant $T_K$, Eq.
(\ref{TK}). For $\nu \gg T_K >T$ the scaling of $j_{LR}$ and
$j_{RR}$ terminates at $D\simeq\nu$ and for $D<\nu$ we have just
one scaling equation, leading to a fixed point at $j_{LL} \to \infty$ and  a Kondo temperature that depends on
$\nu$. However, we will see that 
the tunneling conductance diminishes in the low
temperature regime, $T\ll\nu$. Therefore in the following
discussions we may speak of a single Kondo temperasture,
given by Eq. (\ref{TK}).


\noindent
{\bf Calculations of the Tunneling Conductance in the Weak Coupling Regime}
 are carried out below using the
Keldysh technique in order to treat a system out of equilibrium.
The required quantities to be used elsewhere below are the Keldysh Green's functions (GF)
\begin{equation} \label{gK}
\bar{g}_\alpha=\begin{pmatrix} \bar{g}_\alpha^L&  \bar{g}_\alpha^K \\ 0& \bar{g}_\alpha^A \end{pmatrix},
\end{equation}
where the subscript $\alpha$ refers to the normal metal (L) and TI (R) or the dot (f) electron GF while the superscripts refer to 
retarded (R), advanced (A) and Keldysh (K) types of the GF. For $\alpha=L,R$ the explicit expressions are, 
\begin{eqnarray}
  &&
  \bar{g}_{L}^{R}=-\bar{g}_{L}^{A}=-i\pi\rho_{L},
  \ \ \
  \bar{g}_{L}^{K}(\epsilon)=
  -2i\pi\rho_{L}(1-2f(\epsilon)),
  \nonumber \\
  &&
  \bar{g}_{R}^{R,A}(\epsilon)=
  -\pi\rho_R~\chi(\epsilon\pm i\eta),
  \ \ \
  \\
  &&
  \bar{g}_{R}^{K}(\epsilon)=
  -2i\pi\rho_{R}~
  {\rm{Im}}\chi(\epsilon)~
  \big(1-2f(\epsilon)\big),
  \nonumber
\end{eqnarray}
where  $f(\epsilon)$ is the Fermi  function and 
\begin{equation}   \label{chi-def}
 \chi(z)=\frac{iz}
    {\sqrt{z^2-\nu^2}}~.
    \end{equation}


The current operator for tunnelling  from the NM to the
TI is
\begin{eqnarray}
  I &=&
  \frac{ie}{\hbar}
  \sum_{\sigma\sigma'}
  \Big\{
      \psi_{L\sigma}^{\dag}(0)~
      h_{LR}^{\sigma\sigma'}~
      \psi_{R\sigma'}(0)-
      {\rm{h.c.}}
  \Big\},
  \label{current-def}
\end{eqnarray}
and the differential dc conductance is given by,
$$
  G=
  \frac{\partial{\cal{I}}(V)}
       {\partial{V}},
$$
where the expectation value of
the current is
\begin{eqnarray}
  {\cal{I}}(V) &=&
  \big\langle
      {U_K}^{-1}
      {\cal{T}}
      \big(
          U_K~I(0)
      \big)
  \big\rangle.
  \label{current-Keldysh-def}
\end{eqnarray}
In this expression, ${\cal{T}}$ is the time-ordering operator
and $U_K$ is the evolution operator under $H_K$,
\begin{eqnarray}
  U_K &=&
  {\cal{T}}\exp
  \bigg\{
      -
      \frac{i}{\hbar}~
      \int\limits_{-\infty}^{\infty}dt
      H_K(t)
  \bigg\}.
  \label{S-matrix-def}
\end{eqnarray}

 To lowest (second) non-vanishing order of
perturbation theory in the dimensionless parameters $j_{\alpha
\alpha'}$ and $k_{\alpha\alpha'}=K_{\alpha\alpha'}
\sqrt{\rho_{\alpha}\rho_{\alpha'}}$ , the conductance is
\begin{eqnarray}
  G_2 &=&
  \frac{\pi e^2}{4\hbar}~
  \Big\{
      k_{LR}^2+3j_{LR}^2
  \Big\}
  W(T,V),
  \label{G2}
\end{eqnarray}
where $W(T,V)=w(T,V)+w(T,-V)$,
\begin{eqnarray*}
  w(T,V) &=&
  \frac{1}{4T}
  \int\limits_{\nu}^{\infty}
  \frac{\epsilon~d\epsilon}
       {\sqrt{\epsilon^2-\nu^2}}~
  \frac{1}
       {\cosh^2\big(\frac{\epsilon+eV}{2T}\big)}.
\end{eqnarray*}

The third order correction to the conductance is
\begin{eqnarray}
  G_3 &=&
  \frac{3\pi e^2}{2\hbar}~
  j_{LR}^2
  W(T,V)
  \sum_{\alpha=L,R}
  j_{\alpha\alpha}
  {\cal{L}}_{\alpha}(\bar{D}),
  \nonumber \\
  \label{G3}
\end{eqnarray}
where $\bar{D}$ is a high energy cut off,
\begin{eqnarray*}
  {\cal{L}}_L(\bar{D})
  &=&
  \frac{1}{2}
  \Bigg\{
       \ln
       \bigg(
            \frac{\bar{D}}
                 {\sqrt{(\nu+eV)^2+T^2}}
       \bigg)+
  \nonumber \\ && ~~ +
       \ln
       \bigg(
            \frac{\bar{D}}
                 {\sqrt{(\nu-eV)^2+T^2}}
       \bigg)
  \Bigg\},
  \\
  {\cal{L}}_R(\bar{D})
  &=&
  \ln
  \bigg(
       \frac{\bar{D}}
            {\sqrt{\nu^2+T^2}}
  \bigg).
\end{eqnarray*}
The third order correction $G_3$ contains a
logarithmic term which is, strictly speaking, not small. As a
result, $G_3$ is not small compared with $G_2$ and the expansion
up to terms cubic in $j_{\alpha\alpha'}$ is insufficient. Instead,
we derive the conductance in the leading logarithmic approximation
using the RG equations (\ref{eq-jLL}) -- (\ref{eq-jLR}). 

In the following analysis we split the second
order contribution to the conductance, Eq. (\ref{G2}), in two
parts: one part is due to the exchange cotunneling,'' which is
proportional to $j_{LR}^2$; and one part is due to regular
cotunneling, which is proportional to $k_{LR}^2$. $G_3$, Eq. (\ref{G3}), increases when the
temperature and voltage are lowered, demonstrating the Kondo
anomaly. The regular cotunneling contribution contains $k_{LR}^2$ which does
not grow at low temperatures and bias, and therefore it does not
contribute to the Kondo effect. The exchange cotunneling contains
term $j_{LR}^2$ which demonstrates logarithmic enhancement of the conductance 
at low temperatures [see Eq. (\ref{j(T)-solution})] and contributes to
the Kondo effect. Therefore, we single out the exchange
contribution in the second order term,
\begin{eqnarray}
  G_2^{\rm{exch}}(D) &=&
  \frac{3\pi e^2}{4\hbar}~
  j_{LR}^2(D)
  W(T,V).
  \label{G2-exch}
\end{eqnarray}
 The resulting condition of invariance of the
conductance under the transformation, which corresponds to ``poor
man's scaling'', has the form,
\begin{eqnarray}
  \frac{\partial}{\partial \ln D}
  \bigg\{
       G_2^{\rm{exch}}(D)+
  ~~~~~ ~~~~~
  ~~~~~ ~~~~~
  ~~~~~ ~~~~~
  \nonumber \\ +
       \frac{3\pi e^2}{2\hbar}
       j_{LR}^2
       W(T,V)
       \sum_{\alpha=L,R}
           j_{\alpha\alpha}
           {\cal{L}}_{\alpha}(D)
  \bigg\}
  = 0.
  \label{dG-dD}
\end{eqnarray}
Within the accuracy of this equation, when differentiating the
second term, we should neglect any implicit dependence on $D$
through the couplings $j_{\alpha\alpha'}$. Eq. (\ref{dG-dD}) yields
the scaling equation (\ref{eq-jLR}).

The renormalization procedure should proceed until the bandwidth $D$ is
reduced to a quantity $d(T,V,\nu)=$Mx$[T, eV,\nu]$. 
 At this point, the third order correction to the conductance
is much smaller than $G_2^{\rm{exch}}$ and the current and conductance
can be calculated in the Born approximation, as in Eq. (\ref{G2-exch}).
This situation is similar to the problem considered in Ref.
\cite{Kaminski-Nazarov-Glazman-00}, but the coupling between
the helical modes in the TI introduces an additional energy scale $\nu$
which affects the Kondo physics. When $D$ reduces below $\nu$,
renormalization of both $j_{LR}$ and $j_{RR}$ stops. Then, taking
\begin{eqnarray*}
  d(T,V,\nu) &=&
  \sqrt{\nu^2+(eV)^2+T^2},
\end{eqnarray*}
we get an expression for the conductance for $T,|eV|{\ge}T_K$,
\begin{eqnarray}
  G(T,V) =
  \frac{3\pi^2 G_0}{16}~
  \frac{W(T,V)}
       {\ln^2\big(d(T,V,\nu)/T_K\big)},
  \label{G-low-D}
\end{eqnarray}
where
\begin{eqnarray}
  G_0 &=&
  \frac{e^2}{\pi\hbar}~
  \frac{4 \Gamma_L\Gamma_R}{(\Gamma_L+\Gamma_R)^2}.
  \label{G0-def}
  \nonumber
\end{eqnarray}

\begin{figure}[htb]
\centering
\includegraphics[width=70 mm,angle=0]
   {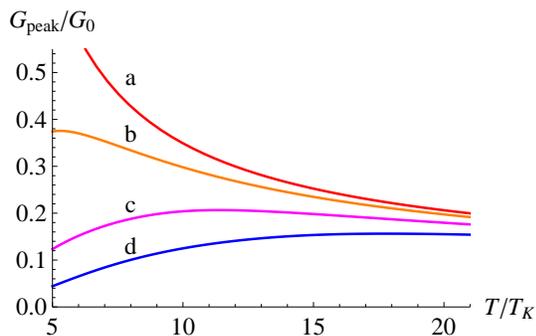}
 \caption{Conductance as function of temperature in the weak coupling regime.
  The curves a, b, c and d correspond to $\nu=0$, $5T_K$,
  $10T_K$ and $15T_K$, respectively. The case $\nu=0$
  corresponds to the standard Kondo effect in NM-QD-NM junction.
  \cite{Kaminski-Nazarov-Glazman-00}}
  \label{Fig-G-weak}
\end{figure}

\begin{figure}[htb]
\centering
\includegraphics[width=70 mm,angle=0]
   {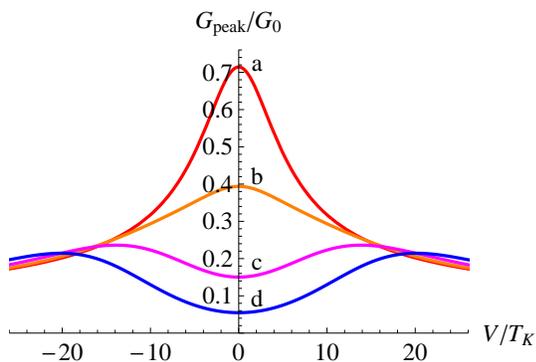}
 \caption{The nonlinear conductance (\ref{G-low-D}) as a function
  of applied bias for $T=5T_K$. The curves a, b, c and d
  correspond to $\nu=0$, $5T_K$, $10T_K$ and
  $15T_K$, respectively}.
  \vspace{-0.2in}
  \label{Fig-GV-weak}
\end{figure}

The total differential conductance is displayed in Fig.
\ref{Fig-G-weak} for $V=0$. It is seen that when the gap in the
spectrum of the TI exceeds $T_K$ (see curves b, c and d),
the conductance increases when the temperature is lowered, reaches
its peak at finite temperature and begins to decrease with further
lowering the temperature.
This occurs because
electron needs an extra energy in order to covercome the gap $\nu$
between the Fermi level of the left lead and the bottom of the
conduction band of the TI. This energy can be transferred from the
internal energy of the electron liquid in the lead and/or from the
voltage difference across the barrier. When the temperature is
lower than the gap, the zero bias conductance decreases 
with temprerature. For comparison, it is noticed that 
the conductance for the
gapless spectrum of the TI (curve a) increases when the
temperature is lowed which corresponds to the standard feature of Kondo
tunneling through the NM-QD-NM junction.
\cite{Kaminski-Nazarov-Glazman-00}

The coupling between the helical states manifests itself also in
the nonlinear tunneling conductance. Fig. \ref{Fig-GV-weak} displays 
the nonlinear conductance as a function of the applied voltage for
$T=5T_K$ and for several values of $\nu$. The zero bias peak
in the conductance corresponding to the standard Kondo tunneling
decreases with the coupling energy $\nu$ (curves a and b) and
splits into two distinct peaks when $\nu$ exceeds the temperature
$T$ (curves c and d). This is one of the central results of the
present paper since it combines the Kondo and the QSH physics.

%
\noindent {\bf Conductance Calculations in the Strong Coupling Regime:} When $T<T_K$,
the mean field slave boson approximation (MFSBA) is used to
calculate the zero bias tunneling conductance. In the limit
$U\to\infty$, the dot can be empty or singly occupied. The dot
operators are written as $d_{\sigma}=b^{\dag}f_{\sigma}$ and
$d_{\sigma}^{\dagger}=f_{\sigma}^{\dag}b$
 where the
slave fermion operators $f_{\sigma}$ and the slave boson operator
$b$ satisfy constraint condition,
\begin{eqnarray*}
 Q &=&
 \sum_{\sigma}
 f_{\sigma}^{\dag}f_{\sigma}+
 b^{\dag}b = 1.
\end{eqnarray*}
This condition is encoded by including a Lagrange multiplier
$\lambda$ in the total action $S$. In the mean field
approximation, we replace the Bose operators $b$ and $b^{\dag}$ by
their expectation values,
\begin{eqnarray*}
  &&
  b \to b_0,
  \ \ \ \ \
  b^{\dag} \to b_0,
  \ \ \ \ \
  b_0 =
  \sqrt{\langle{b^{\dag}b}\rangle}.
\end{eqnarray*}
At the mean field level the constraint condition is satisfied only
on the average.

The partition function $Z(\alpha)$ is calculated by integrating
the action over slave fermion field. Here the source field
$\alpha$ is coupled to the current operator,
\begin{eqnarray}
  I &=&
  \frac{iet_L}{\hbar}
  \sum_{\sigma}
  \Big[
      \psi_{L\sigma}^{\dag}(0)
      b^{\dag}f_{\sigma}-
      {\rm{h.c.}}
  \Big].
  \label{I-def}
\end{eqnarray}
The effective action in the MFSBA is Gaussian and depends on two
real numbers, the boson field $b_0$ and the chemical potential
(Lagrange multiplier) $\lambda$. Integrating the action leads to
the partition function,
\begin{eqnarray}
  \ln Z(\alpha) =
  -{\rm{tr}}~\ln
  \Big\{
      G_{f}^{-1}-
      \frac{e \alpha_q t_L^2 b_0^2}{\hbar}~
      \big[
          \bar{g}_L,
          \tau_x
      \big]
  \Big\},
  \label{Z-dot}
\end{eqnarray}
where
\begin{eqnarray}
  G_{f}^{-1} &=&
  g_{f}^{-1}-
  t_L^2 b_0^2 \bar{g}_L-
  t_R^2 b_0^2 \bar{g}_{R}.
  \label{G-f}
\end{eqnarray}
Here $g_f$ is the GF of the (non-interacting) electron in the QD with shifted energy
level, $\epsilon_0\to\bar\epsilon_0=\epsilon_0+\lambda$,
\begin{eqnarray*}
  g_{f}^{R/A}(\epsilon) =
  \frac{1}{\epsilon-\bar\epsilon_0\pm{i\eta}},
  \ \ \ \ \
  g_{f}^{K}(\epsilon) =
  -\frac{2i\eta(1-2f(\epsilon))}
        {(\epsilon-\bar\epsilon_0)^2+\eta^2},
\end{eqnarray*}

\begin{figure}[htb]
\centering
\includegraphics[width=80 mm,angle=0]{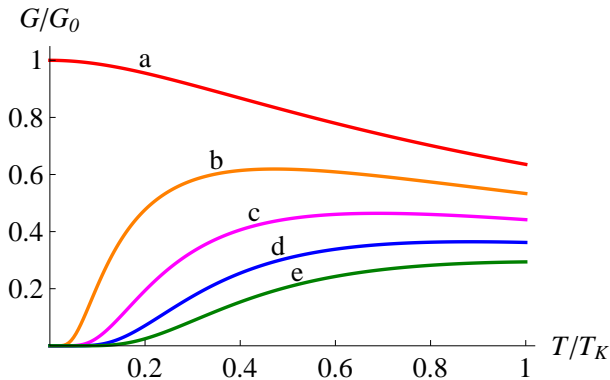}
 \caption{The zero bias conductance as a function of temperature
   in the strong coupling limit ($T<T_K$). The curves a, b, c, d
   and e correspond to $\nu=0$, $0.2T_K$, $0.4T_K$, $0.6T_K$ and
   $0.8T_K$, respectively. Here we take $\Gamma_L=\Gamma_R$.}
   \vspace{-0.in}
\label{Fig-G-strong}
\end{figure}

The MFSBA is reliable in equilibrium, $V=0$. Therefore we will
consider below the temperature dependence of the zero bias
conductance. In equilibrium, the mean field solutions for $b_0$
and $\lambda$ minimize the free energy,
$$
  F=
  -T
  \sum_{\omega_n}
  {\rm{tr}}~
  \ln
  G_f^{-1}(i\omega_n)+
  \lambda b_0^2,
$$
where the last term is the slave boson kinetic part of the free
energy due to the constraint, $G_f^{-1}(i\omega_n)$ is the
Matsubara's GF. The mean field equations are
\begin{eqnarray}
  b_0^2 =
  \frac{2}{\pi}
  \arctan
  \bigg(
       \frac{2(\epsilon_0+\lambda)}
            {b_0^2 \Gamma}
  \bigg),
  \ \ \
  \lambda =
  \frac{\Gamma}{\pi}
  \ln
  \bigg(
       \frac{\pi\bar{D}}
            {b_0^2 \Gamma}
  \bigg).
  \label{eqs-MFSBA}
\end{eqnarray}
These equations are solved for $b_0$ and $\lambda$ with the solutions,
\begin{eqnarray}
  b_0^2=
  \frac{\pi T_K}{\Gamma},
  \ \ \ \ \
  \epsilon_0+\lambda=
  \frac{\pi^3 \big(T_K\big)^2}{2\Gamma},
  \label{sol-MFSBA}
\end{eqnarray}
where $T_K$ is the Kondo temperature given by Eq.
(\ref{TK}). Then the linear conductance for $T<T_K$ is
\begin{eqnarray}
  G =
  \frac{\pi T_K G_0}{4T}
  \int
  \frac{d\epsilon~
        {\cal{S}}_f(\epsilon)}
       {\displaystyle
        \cosh^2
        \Big(
            \frac{\epsilon}{2T}
        \Big)}~
  \frac{(\Gamma_L+\Gamma_R)
        {\rm{Im}}\chi(\epsilon)}
       {\Gamma_L+\Gamma_R~
        {\rm{Im}}\chi(\epsilon)},
  \label{G-int}
\end{eqnarray}
where
\begin{eqnarray*}
  &&
  {\cal{S}}_f(\epsilon)=
  \frac{T_K\gamma(\epsilon)}
       {(\epsilon-\bar\epsilon_0)^2+
        \big(T_K\big)^2\gamma^2(\epsilon)},
  \\
  &&
  \gamma(\epsilon)=
  \frac{\pi}{2}~
  \frac{\Gamma_L+\Gamma_R~{\rm{Im}}\chi(\epsilon)}
       {\Gamma_L+\Gamma_R},
\end{eqnarray*}
$G_0$ is given by Eq. (\ref{G0-def}), $\chi(\epsilon)$ is given by
Eq. (\ref{chi-def}).

The zero bias conductance as a function of temperature is shown in
Fig. \ref{Fig-G-strong} for $\nu=0$, $0.2T_K$,
$0.4T_K$, $0.6T_K$ and $0.8T_K$. The temperature
dependence of the conductance is similar to the one shown in Fig.
\ref{Fig-G-weak} for $T>T_K$. In both cases, when $T>\nu$, a
peak in the conductance occurs (see curves b, c, d and e). This
peak decreases with $\nu$ and shifts towards high temperatures.
For $T<\nu$, the conductance decreases with decreasing in $T$ and
vanishes when $T\to0$. This is a manifestation of the coupling
between the QSH states on the opposite sides of the TI. For
$\nu=0$, the conductance has a usual peak at $T=0$.
%

\noindent {\bf Summary:} The linear and non-linear conductance in
a system consisting of a quantum dot (tuned to the Kondo regime)
connected to a metal lead on the left side and to a topological
insulator on the right side is evaluated using Keldysh technique.
Renormalization group analysis is performed in the weak coupling
regime ($T \gg T_K$) while the MFSBA is used at at the
strong coupling regime $T<T_K$. When the coupling energy
$\nu$ between QSH helical states exceeds the temperature, the differential
conductance develops a dip at low temperature, in contrast with
the standard zero bias anomaly Kondo peak prevailing in QD
connected to metallic leads. Our analysis then shows that the
Kondo resonance is very sensitive to edge state coupling and can
be used to probe the features of QSH helical modes.\\
{\bf Acknowledgment:} We thank the Israeli Science Foundation for supporting our research under grant 1173/08.



\begin{thebibliography}{99}

\bibitem{Teo-Kane-10} J.C.Y. Teo, C.L. Kane, Phys.Rev. {\bf B 82},
         115120 (2010).

\bibitem{3DTI-08} A.P. Schnyder, S. Ryu, A. Furusaki, and
         A.W.W. Ludwig, Phys. Rev. B 78, 195125 (2008).

\bibitem{TI-and-TS-09} A.P. Schnyder, S. Ryu, A. Furusaki, and
         A.W.W. Ludwig, AIP Conf. Proc. 1134, 10 (2009).

\bibitem{Kane-Mele-prl03} C.L. Kane and E.J. Mele,
         Phys. Rev. Lett. {\bf{95}}, 146802 (2005).

\bibitem{Liang-Kane-Mele-prl07} Liang Fu, C.L. Kane, and
         E.J. Mele, Phys. Rev. Lett. {\bf{98}}, 106803 (2007).

\bibitem{Hasan-Kane-rmp10} M.Z. Hasan and C.L. Kane,
         Rev. Mod. Phys. {\bf{82}}, 3045 (2010).

\bibitem{Liang-Kane-prb07} Liang Fu and C.L. Kane,
         Phys. Rev. B {\bf{76}}, 045302 (2007).

\bibitem{Liang-Kane-prl08} Liang Fu and C. L. Kane,
         Phys. Rev. Lett. {\bf{100}}, 096407 (2008).

\bibitem{Bi2Te3-and-Sb2Te3-prl09} D. Hsieh, Y. Xia, D. Qian,
         L. Wray, F. Meier, J.H. Dil, J. Osterwalder, L. Patthey,
         A.V. Fedorov, H. Lin, A. Bansil, D. Grauer, Y.S. Hor,
         R.J. Cava, and M.Z. Hasan, Phys. Rev. Lett. {\bf{103}},
         146401 (2009).

\bibitem{Bi2Te3-prl09} T. Zhang, P. Cheng, X. Chen, J.-F. Jia,
         X. Ma, K. He, L. Wang, H. Zhang, X. Dai, Z. Fang, X. Xie,
         and Q.-K. Xue, Phys. Rev. Lett. {\bf{103}}, 266803 (2009)

\bibitem{Bi2Se3-prb09} Y.S. Hor, A. Richardella, P. Roushan,
         Y. Xia, J.G. Checkelsky, A. Yazdani, M.Z. Hasan,
         N.P. Ong, and R.J. Cava, Phys. Rev. B {\bf{79}}, 195208 (2009).

\bibitem{HgMnTe-TI-08} C.-X. Liu, X.-L. Qi, X. Dai, Z. Fang, and
         S.-C. Zhang, Phys. Rev. Lett. {\bf{101}}, 146802 (2008).

\bibitem{TI-10} N.H. Lindner, G. Refael and V. Galitski,
         Nature Physics {\bf{7}}, 490 (2011).

\bibitem{Bi2Se3-photo-prl09} J.G. Checkelsky, Y.S. Hor, M.-H. Liu,
         D.-X. Qu, R.J. Cava, and N.P. Ong, Phys. Rev. Lett. {\bf{103}},
         246601 (2009).

\bibitem{konig} M. Konig et al., Science {\bf{318}}, 766, (2007).

\bibitem{egger} R. Egger, A. Zazunov, and A. Levy Yeyati,
         Phys. Rev. Lett. {\bf{105}}, 136403 (2010).

\bibitem{Tkachev-11} G. Tkachov, E. M. Hankiewicz, Phys. Rev. B
        {\bf{83}}, 155412 (2011).

\bibitem{TI-trans-09} Y. Tanaka, T. Yokoyama, and N. Nagaosa,
         Phys. Rev. Lett. {\bf{103}}, 107002 (2009).

\bibitem{TI-trans-10} H.-Z. Lu, W.-Y. Shan, W. Yao, Q. Niu, and
         S.-Q. Shen, Phys. Rev. B {\bf{81}}, 115407 (2010).

\bibitem{TI-trans-11} C. Timm, {\tt arXiv:1111.2245}.

\bibitem{GKY-11} A. Golub, I. Kuzmenko, and Y. Avishai,
         Phys. Rev. Lett. {\bf{107}}, 176802 (2011).

\bibitem{ScanTunMicr-prl11} Y. Okada, C. Dhital, Wen-Wen Zhou,
         Hsin Lin, S. Basak, A. Bansil, Y. -B. Huang, H. Ding,
         Z. Wang, Stephen D. Wilson, V. Madhavan, Phys. Rev. Lett.,
         {\bf{106}}, 206805 (2011).

\bibitem{TI-tun-10} C.-Y. Seng, T.-K. Ng, {\tt arXiv 1012.5867}.

\bibitem{Schrieffer-Wolff-66} J.R. Schrieffer and P.A. Wolff,
         Phys. Rev. {\bf{149}}, 491 (1966).

\bibitem{PustGlazm00}{\bf{25}} M. Pustilnik and L. Glazman, Phys. Rev. Lett.
        {\bf{85}}, 2993 (2000); {\tt cond-mat/0102458}.

\bibitem{KikoinAvishai02}{\bf{26}} K. Kikoin, Y. Avishai, Phys. Rev. B
        {\bf{65}}, 115329 (2002); {\tt cond-mat/0107473}.

\bibitem{Kaminski-Nazarov-Glazman-00} A. Kaminski, Yu. V. Nazarov,
         and L.I. Glazman, Phys. Rev. B {\bf{62}}, 8154 (2000).

\end{thebibliography}
\end{document}